\documentclass[12pt,reqno]{amsart}
\usepackage{graphicx,color}
\usepackage[centertags]{amsmath}
\usepackage{enumerate}

\begin{document}

\title[The Third Way Of Probability]{The Third Way Of Probability \& Statistics: Beyond Testing and Estimation To Importance, Relevance, and Skill}

\author{William M. Briggs}
\email{matt@wmbriggs.com}
\urladdr{wmbriggs.com} 

\begin{abstract}
There is a third way of implementing probability models and practicing. This is to answer questions put in terms of observables. This eliminates frequentist hypothesis testing and Bayes factors and it also eliminates parameter estimation.  The Third Way is the logical probability approach, which is to make statements $\Pr(Y \in y | X,D,M)$ about observables of interest $Y$ taking values $y$, given probative data $X$, past observations (when present) $D$ and some model (possibly deduced) $M$. Significance and the false idea that probability models show causality are no more, and in their place are importance and relevance. Models are built keeping on information that is relevant and important to a decision maker (and not a statistician). All models are stated in publicly verifiable fashion, as predictions. All models must undergo a verification process before any trust is put into them.  
\end{abstract}

\maketitle

\section{Introduction}

Testing and estimation, whether in the frequentist or Bayesian implementations, form the backbone of classical analyses. This is unfortunate because the questions answered by testing and estimation are not of interest to anyone except statisticians.  A typical question would be, ``How is age related to the uncertainty in contracting cancer of the albondigas?" The tactic is to say whether the p-value in a parameterized probability model relating age and cancer was wee, or to say that a Bayes factor for the same parameter has a certain value, or by stating an estimate of that unobservable parameter. These are obviously not answers to the question.  Error is introduced, and on a breathtaking scale, by supposing the classical statements {\it are} the answer. This happens in almost all published instances of statistical analysis.

Disputes in statistics center around Bayesian versus frequentist methods. While there is a wide chasm between the interpretation of probability in these two philosophies, the practical differences are minor. Both methods devote themselves to parameters. Yet parameters are of no direct interest in answering questions about the uncertainty of observables (such as cancer of the albondigas). However, a doleful tradition has developed which says they are. If a p-value is wee, or a Bayes factor large, or a parameter estimate a certain value, uncertainty in the observable is thought to be adequately quantified. But this is not so: no matter how much certainty is present in parameters, the uncertainty in the observable must always be greater. Therefore relying on parameter-certainty as a proxy for observable-certainty is to be over-confident.

Some have noticed this discrepancy, such as Seymour Geisser, see \cite{Gei1993,Geisser1984}, but knowledge of his work is limited, hampered, no doubt, by lacking a well-articulated replacement for the classical methods. The Third Way proposed below is (it is hoped) such a replacement.  To maintain consistency the interpretation of philosophy is as logic is adopted. This is expounded in, among others, \cite{Jay2003,Cox1961,Key2004,Lap1996} as well as philosophically  in \cite{CamFra2004,Fra2001,Fra2001b}. Those familiar with objective Bayes will recognize many mathematical parallels, e.g. \cite{Wil2010} (although this author does not cross over into full logical probability).  The only piece of philosophy necessary to grasp the basics of logical probability is to recognize that {\it all} probability is conditional. That is, there is no such thing as unconditional probability of any proposition.\footnote{The skeptical reader is invited to try to discover the probability of a proposition that relies on no evidence.}

The gist of the Third Way is simple: answer the questions put to us about the uncertainty in observables. Eliminate all talk of parameters, all notion of testing, and all manner of estimation.  If somebody asks, ``How is age related to the uncertainty in contracting cancer of the albondigas?", we tell them.

\section{The Third Way}

Let past observables be labeled $D = (Y,X)_{\mbox{old}}$, where $Y$ is the observable in which we want to quantify or explain our uncertainty, and $X$ are the premises or observables assumed probative of $Y$ (the dimensions of each will be obvious in context).  Let the premises which lead to a probability model (if one is present) be labeled $M$. And let $X = X_{\mbox{new}}$ be the premises or assumed values of new observables. The goal of all probability modeling is this:
\begin{equation}
  \label{third}
  \Pr(Y \in y | X,D,M),
\end{equation}
where $y$ are values of the observable $Y$ which are of interest to some decision maker.  Models should be rare, because most probability is not quantifiable---and we must resist the temptation to force quantification by making up scientific-sounding numbers.  But even if we do, (\ref{third}) can be calculated, as long as supply the premises which led to our creations.\footnote{Readers are asked to express their agreement with this proposition on a scale of -1.74 to $e^{\pi}$.} 

Although it is obvious, the equation reads, ``The probability $Y$ takes the values $y$ given the premises or assumptions $X$, the past data $D$, and the model $M$." If the model is parameterized and Bayesian philosophy is adopted, (\ref{third}) is the posterior predictive distribution, and $M$ incorporates those premises or assumptions from which the priors are deduced (see e.g. \cite{BerSmi2000}). If a frequentist philosophy is adopted, there are many difficulties and inconsistencies in interpretation, but I will not discuss them here; the meaning of equation is plain enough. The key is that no parameters are explicit in (\ref{third}); the uncertainty in them has been ``integrated out." Only observables and plain assumptions remain. Logical probability would supply premises from which the model $M$ is deduced (there would be no parameters thus no priors).

Equation (\ref{third}) eliminates, or rather combines, the efforts of testing and estimation into one form.  The focus is entirely on observables and the assumptions made and their effect on the uncertainty of not-yet-seen or unknown values of $Y$. Not-yet-seen values of $Y$ are those unknown or assumed unknown; usually they are as yet unmeasured, e.g. in the future.

A simple example is a die roll in which $M$ = ``This is a six-sided object with labels one through six and which when tossed must show only one side." The model is deduced based on these premises. There is no $X$ probative beyond $M$ and $D$ can be absent or can be a record of previous flips (i.e. $X$ and $D$ are null or are assumed not probative). An application of (\ref{third}) is $\Pr(Y = 6 | X,D,M) = 1/6$.  Because the model was deduced, no parameters were ever present. 

It is unfortunately rare that models are deduced; most are posited {\it ad hoc}. That is, $M$ is usually ``I'm using regression", i.e. an act of will. Model deduction can be accomplished if the measurement of observable are properly accounted for. See Briggs \cite{Bri2016} for details.  Since that problem takes us too far afield, I'll stick with arbitrary or capricious models.  Suppose we are interested in $Y$ = ``First-year college grade point average" of students. Observations $X$ thought probative are the high school grade point average and SAT score. The model $M$ will be ordinary regression.  The goal is to produce statements like this:
\begin{equation}
  \label{cgpa}
  \Pr(Y > 3.8 | X_{\mbox{h}} = 3.5, X_{\mbox{s}} = 1160 ,D,M),
\end{equation}
where the subscript ``h" is for high school GPA, and ``s" is for SAT. The $D$ are past observations.  Since regression uses continuous normal probability, we unfortunately cannot ask about observables like ``$Y = 4.0$" and must restrict our attention to intervals.  

At any rate, the main questions of interest are only two: (1) how do changing values of $X_{\mbox{h}}$ and $X_{\mbox{s}}$ change the uncertainty in $Y$ in the presence of $D$, $M$, and $y$, and (2) of what value or descriptive power is $M$?  $M$ includes $X_{\mbox{h}}$ and $X_{\mbox{s}}$ as components.

There is no notion of ``significance" in either of these questions, and no notion of correctness in any instance where the model was not deduced. All probability is conditional, and that means the probabilities given in equations like \ref{cgpa} are correct.  If, at some later point, it is decided that $X_{\mbox{s}}$ is of no interest and the model is updated to reflect this, then the probabilities derived from the new model are also correct. One model may be superior to another, however, but only with respect to decisions made conditional on the models. 

There is more work to be done by model builders in the Third Way because values of $y$ must also be chosen, and so do values of $X$. This implies a model may be useful in some decision contexts and of no use or even harmful in others.  There is and should be no default or automatic levels of usefulness. The last thing the field of probability and statistics needs is another magic number {\it a la} ``significance" with p-values. 

The importance of probative observables and assumptions $X$ is thus also a matter of decision and cannot be made automatically. This is plain from equation (\ref{third}), which encapsulates {\it all} we know about $Y$ given our assumptions. It is we who made these assumptions, and we who can change them.  Again, the only ultimately true model of $Y$ is that which is deduced from true premises, as in the die example. Every other model is therefore only useful or not conditional on the premises we assume. Since most models are {\it ad hoc}, as regression always is, we can only speak of usefulness. Deduced models are true by definition and thus nothing more need be done with them except make predictions. Deduced models do not even need to be verified.

The idea in the Third Way is, conditional on $D$ and $M$, to vary $X$ in the range of expected, decisionable, or important values to some decision maker and see how these change the probability of $Y \in y$. If a particular $X$ as it ranges along the values we choose do not change the probabilities of $Y \in y$ in any important way, then these $X$ are themselves not important. The opposite is also true. Importance is a matter of decision, which varies by decision maker. Importance is not a probability or statistical concept and therefore cannot be ascertained within probability models.  If the probability of $Y \in y$ changes in any as $X$ does, then that $X$ is relevant to understanding $Y$, else it is not. Relevance is a probabilistic concept, but as the reader will see, it is almost always present given the assumption that the $X$ is causally related to the $Y$.  If $X$ is known not be causally related to $Y$, then $X$ is irrelevant by definition.

There is no hypothesis testing in the frequentist or Bayesian sense (as implied by Bayes factors, for instance). And there is no estimation of parameters. There are only plain, understandable, and verifiable probability statements. These probability statements can and should and must be verified. The Third Way allows communication of model goodness and usefulness in an intelligible, actionable manner.  It reduces over-certainty but cannot eliminate it unless models are deduced.

\section{Example}

There are 100 observations of first-year college students' grade point averages. We want to quantify the uncertainty in the GPAs of new students given these observations, and also given information thought probative, in this case high school GPAs and SAT scores.\footnote{The data is available at http:\\\\wmbriggs.com\\public\\sat.txt.} We assume an {\it ad hoc} ordinary regression model.  If we adopt the Bayesian philosophy, we need priors, and here an assumption of ``flat" priors will do. As is well known, in ordinary regression this assumption matches the answers given by frequentist philosophy. But it doesn't matter. Any premises that give different priors will do. Our purpose here is not (directly) investigating priors, but the uncertainty inherent in $Y$ given the assumptions we make. 

It turns out
\begin{equation}
  \label{cgpa2}
  \Pr(Y > 3.8 | X_{\mbox{h}} = 3.5, X_{\mbox{s}} = 1160 ,D,M_{\mbox{h,s}}) = 0.038.
\end{equation}
\noindent Notice that the dependence of the model on the assumptions has been annotated, as it {\it always should be}. If, given $D$, we insist on $M$ and on the presence of  $X_{\mbox{h}}$ and $X_{\mbox{s}}$, then this is the final and true answer. Nothing more need be done. The values picked for $y$, $X_{\mbox{h}}$, and $X_{\mbox{s}}$ are those I, and perhaps nobody else, thought important. A different decision maker might pick different values. 

But suppose I am interested in the relevance of $X_{\mbox{h}}$. Its presence is an assumption, a premise, one that I felt important to make. There are several things that can be done. The first is to remove it. That leads to 
\begin{equation}
  \label{cgpa3}
  \Pr(Y > 3.8 | X_{\mbox{s}} = 1160 ,D,M_{\mbox{s}}) = 0.0075.
\end{equation}
\noindent Notice first that {\it both} (\ref{cgpa2}) and (\ref{cgpa3}) are {\it correct}. The probability in (\ref{cgpa2}) is 5 times larger than in (\ref{cgpa3}). This is a measure of relevance and importance, given $y = 3.8$ and $X_{\mbox{s}} = 1160$. Importance and relevance, like probability itself, are always conditional on our assumptions.  A second measure of importance is the change in probabilities when  $X_{\mbox{h}}$ is varied. That can be seen in the following figure.

\begin{figure}[tb]
    \begin{center}
        \scalebox{.6}{\includegraphics{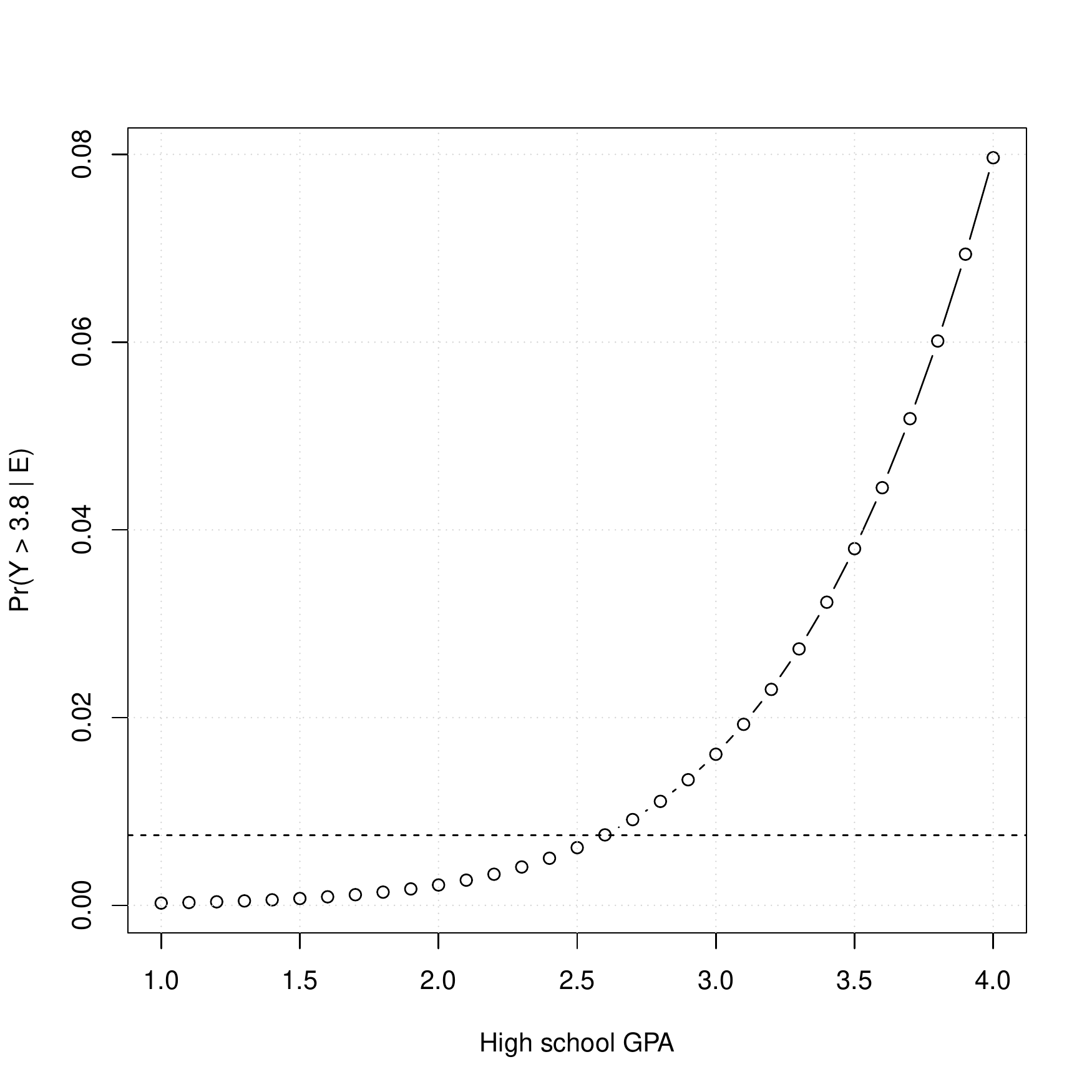}}
    \end{center}
\caption{\label{fig1} $\Pr(Y > 3.8 | X_{\mbox{h}} = x, X_{\mbox{s}} = 1160 ,D,M_{\mbox{h,s}})$ allowing $X_{\mbox{h}}$ to vary from 1 to 4 in increments of 0.1. The notation on the figure conditions on E, which is shorthand for all the evidence we have. The dashed line is $\Pr(Y > 3.8 | X_{\mbox{s}} = 1160 ,D,M_{\mbox{s}})$. }
\end{figure}

There is a change from about 0 to 8\% over the range of high school GPAs. If high school GPA was not probative of $Y > 3.8$ given these premises then the graph would be flat, indicating no change. In other words, it would resemble the dashed line, which is (\ref{cgpa3}), the model without high school GPA.  Is this ``departure" from flatness important?  There is no single answer to this question. That entirely depends on the uses to which this model is put. If a decision would be made differently given these varying values of the probability, then high school GPA is important, otherwise it is not. The answer is not a matter of probability or statistics.  That the line is not flat is {\it proof}, however, that, given $M$, $y$, and $X_{\mbox{s}}$, knowledge of high school GPA is {\it relevant} to knowledge of $Y$. 

It cannot be emphasized too strongly that importance and relevance are conditional, just as probability is. A linear function of high school GPA added to a regression model already supplied with high school GPA would be irrelevant. It might be that high school GPA is relevant or important at some levels of SAT, and irrelevant and unimportant at others, and the same is true for the goodness measures of SAT. Information $X$ is not isolated and is related to all the assumptions we made, and these include the other $X$ in the model.

Now add the information $X_{\mbox{w}}$ = ``hours studied a week" to the model and create a relevance plot for it.

\begin{figure}[tb]
    \begin{center}
        \scalebox{.6}{\includegraphics{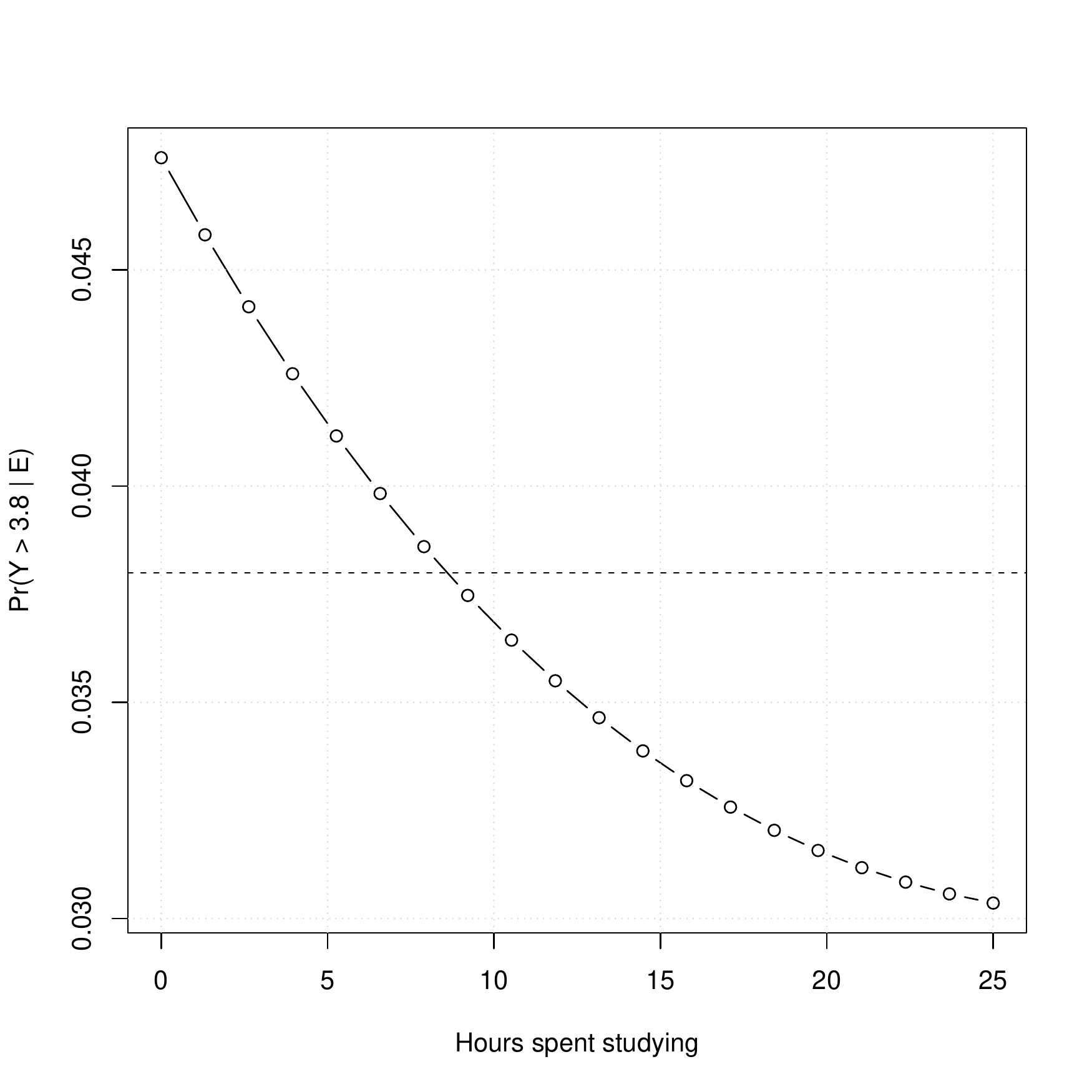}}
    \end{center}
\caption{\label{fig2} $\Pr(Y > 3.8 | X_{\mbox{h}} = 3.5, X_{\mbox{s}} = 1160, X_{\mbox{w}} = x ,D,M_{\mbox{h,s,w}})$ allowing $X_{\mbox{w}}$ to vary from 0 to 26. The notation on the figure conditions on E, which is shorthand for all the evidence we have. The dashed line is $\Pr(Y > 3.8 | X_{\mbox{h}} = 3.5, X_{\mbox{s}} = 1160 ,D,M_{\mbox{h,s}})$. }
\end{figure}

The (conditional, as always) probability of $Y > 3.8$ varies little, from 0.038 to about 0.044, a change of only 0.006. Time studying is relevant because the probability differs from the straight line, which is the probability using the model {\it sans} time studying. But it is important? Given the values of high school GPA, SAT, and $y$, the change from 0.038 to 0.044 is small and therefore many decision makers might conclude that adding time studying provides no benefit to our understanding. Of course, such a small change might be useful to somebody. It is not the statistician's job to decide, but the decision maker's. 

I cheated, because ``time studying" isn't that at all. In fact, it is made-up numbers using R's rnorm() and abs() functions.  It is therefore not surprising that the pertinent probability should vary little. That it does at all is only because of coincidence. By ``coincidence" I do not mean ``randomness" or ``chance", because I know the determinative cause of these numbers, but happenstance, a known lack of causal connection between the made-up numbers and the observable of interest.  We should add information presumed probative into a model {\it only} if we have a plausible belief that the information is related (somehow) to the cause of the observable of interest. If this connection is lacking, the information should not be added. Thus ``time studying" should certainly be removed. Understanding causality in probability models is difficult subject beyond the scope of this paper, except to emphasize that probability models cannot deduce cause (though because of hypothesis testing just about everybody thinks they can; see Briggs \cite{Bri2015} for more details).

The Third Way strategy is to create scenarios that are of direct interest to a decision maker, the person or persons who will use the model. Plots like those above can be made at the values of the probative observables in which the decision maker is interested. There are no one set of right or proper values, except in the trivial sense of excluding values that are, given exterior information, known to be impossible.  For instance, given our knowledge of grade points, the value $X_{\mbox{h}} = -17$ is impossible.  Assessing relevance and importance for large models will not be easy. But who insisted it should be?  That classical statistical procedures now make analysis so simple is part of the problem we're trying to correct. 

Given the model and old observations, every set of $X_{\mbox{h}}$ and $X_{\mbox{s}}$, at some $y$, produce a prediction.  For instance, for future students with $X_{\mbox{h}} = 3.5$ and $X_{\mbox{s}} = 1200$ the (conditional) probability that $Y>3.8$ is 0.045.  In regression, incidentally, we do not have to restrict ourselves to a fixed $y$, because the model will produce a prediction of every possible value of $Y$. These predictions can and {\it must} be verified. An example of such a report is given in Table \ref{Table1}. Considerable art and thinking will have to go into presenting predictions from a model loaded with probative $X$. This may seem like a drawback, but in fact it is a boon. Far too many models are crammed with extraneous ``controlling variables" the usefulness is scarcely ever considered. This approach forces such consideration and encourages leanness; which is to say, models without fat. Lean models are not necessarily small.

\begin{table}
\begin{center}
\caption{The predictions for $\Pr(Y<2 | X_{\mbox{h}} = h, X_{\mbox{s}} = s,D,M)$ and $\Pr(Y>3 | X_{\mbox{h}} = h, X_{\mbox{s}} = s,D,M)$ for common values of $h$ and $s$, and two points of $y$ thought to be of interest. Every $X$ in the model appears in the Table.\label{Table1}}
\begin{tabular}{lccccccccc}
 & \multicolumn{4}{c}{$\Pr(Y<2|h,s,D,M)$} & &  \multicolumn{4}{c}{$\Pr(Y>3|h,s,D,M)$}\\
 $h$/$s$ & 400  & 800  & 1200  & 1600   & & 400     &    800  & 1200   & 1600\\\hline
 0.5 & 0.99 & 0.94 & 0.760 & 0.4600 & & 3.9e-05 & 0.00065 & 0.0093 & 0.071\\
 1.0 & 0.98 & 0.89 & 0.640 & 0.3400 & & 1.4e-04 & 0.00180 & 0.0200 & 0.120\\
 1.5 & 0.95 & 0.81 & 0.510 & 0.2200 & & 5.0e-04 & 0.00500 & 0.0420 & 0.190\\
 2.0 & 0.90 & 0.70 & 0.370 & 0.1300 & & 1.7e-03 & 0.01300 & 0.0820 & 0.290\\
 2.5 & 0.83 & 0.57 & 0.250 & 0.0730 & & 5.0e-03 & 0.03100 & 0.1500 & 0.420\\
 3.0 & 0.73 & 0.43 & 0.160 & 0.0360 & & 1.4e-02 & 0.06600 & 0.2400 & 0.550\\
 3.5 & 0.61 & 0.30 & 0.089 & 0.0170 & & 3.3e-02 & 0.13000 & 0.3700 & 0.680\\
 4.0 &0.48 & 0.20 & 0.048 & 0.0074 & & 6.9e-02 & 0.22000 & 0.5000 & 0.790\\
\end{tabular}
\end{center}
\end{table}

The verification strategy is this. A model is built using importance and relevance, as above. It is then released into the wild, as it were, to wait for new data to arise. Every new data point will have a value of $(X_{\mbox{h}}, X_{\mbox{s}}, Y)$. These are fed into the model and the prediction for the probability of $Y$ is given (with, say, $y = Y$).  These prediction-observable pairs are then evaluated in relation to a proper score and possibly also with respect to a simpler version of the model, in a move to assess (what is called) model skill. This is called the verification process. It is what happens naturally in, say, engineering and physics, where models are forced to meet reality. See, {\it inter alia}, \cite{GneRaf2007,Mur1991,MurWin1987,BriZar2007} for details on verification, an incredibly broad and rich subject. All that can be done here is to give a hint about the best tools. At bottom, the best verification is to assess how well the model performed in the decisions made with it. 

Probability leakage is also discoverable \cite{Bri2013}. This is when the model gives non-zero probability to values of $Y$ which are known to be impossible given external evidence, evidence which is usually ignored in the model-building process (regression, for instance, is horribly over-used). For instance, we learn that at this college a GPA of 4 is the maximum. Yet in our model with $(X_{\mbox{h}} = 4$ and $X_{\mbox{s}} = 1400$ the probability that $Y>4$ is 0.105. This is substantial leakage and a guarantee of model weakness. 

We can also compare our touted model with a simpler model, which is perhaps a standing competitor or is otherwise natural to consider. In the example, time spent studying was revealed to be faked. But suppose I were to give the data to a statistician and not tell him of the fraud. Time studying sounds plausibly causally related to grade point. The check of relevance and importance do not excite. Thus it is reasonable to say two models are in contention, $M_{\mbox{h,s,w}}$ and $M_{\mbox{h,s}}$. If, considering whatever proper score we are using, $M_{\mbox{h,s,w}}$ cannot ``beat" $M_{\mbox{h,s}}$, $M_{\mbox{h,s,w}}$ is said not to have skill in relation to $M_{\mbox{h,s}}$.  This technique is widely used in meteorology (see the references above). In regression, the so-called null model (only with an ``intercept") is always available as a comparator.

A popular (but by no means perfect) proper score is the continuous rank probability score (CRPS), which is the squared distance between the cumulative distribution function of each prediction and the staircase step-function empirical cumulative distribution of each observable \cite{GneRaf2007}. Smaller is better. The mean or summed CRPS is presented over multiple observations, but it can also be examined for each observation. Skill is a relative measure of improvement in score, usually $(\mbox{score}_{\mbox{full}} - \mbox{score}_{\mbox{partial}})/\mbox{score}_{\mbox{partial}}$, comparing a full or larger model with a partial, less complex, or otherwise natural comparator. A perfect skill is 1; values less than 0 are not skillful. Models without skill should not be used. Skill, as is obvious, depends on the score.

The mean CRPS for the ``full" model with ``time studying" is 0.0734, while for the ``partial" model without it is 0.0749. The skill score is 0.02. This shows the full model is superior, but only just barely.  Here is a per-observation analysis of CRPS and skill. 
\begin{figure}[tb]
    \begin{center}
        \scalebox{.6}{\includegraphics{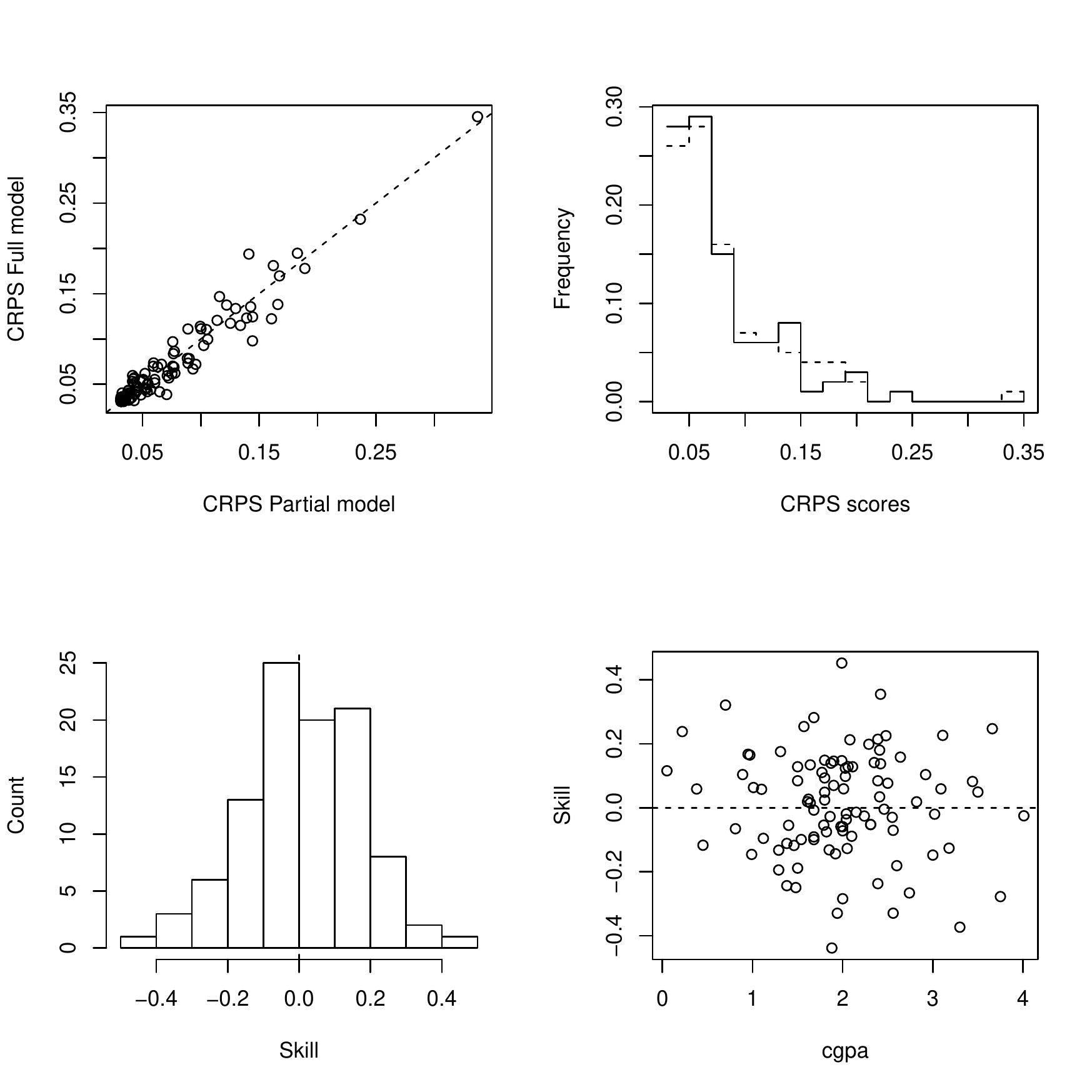}}
    \end{center}
\caption{\label{fig3} Top left: the CRPS calculated for each observation for the full and partial models; a one-to-one line is over-plotted. Top right: the histograms of CRPS scores; the partial model is the dashed line. Bottom left: the skill score for each observation. Bottom right: The skill plotted for each observation of college grade point. Skill is had for values greater than 0.}
\end{figure}

\begin{figure}[tb]
    \begin{center}
        \scalebox{.6}{\includegraphics{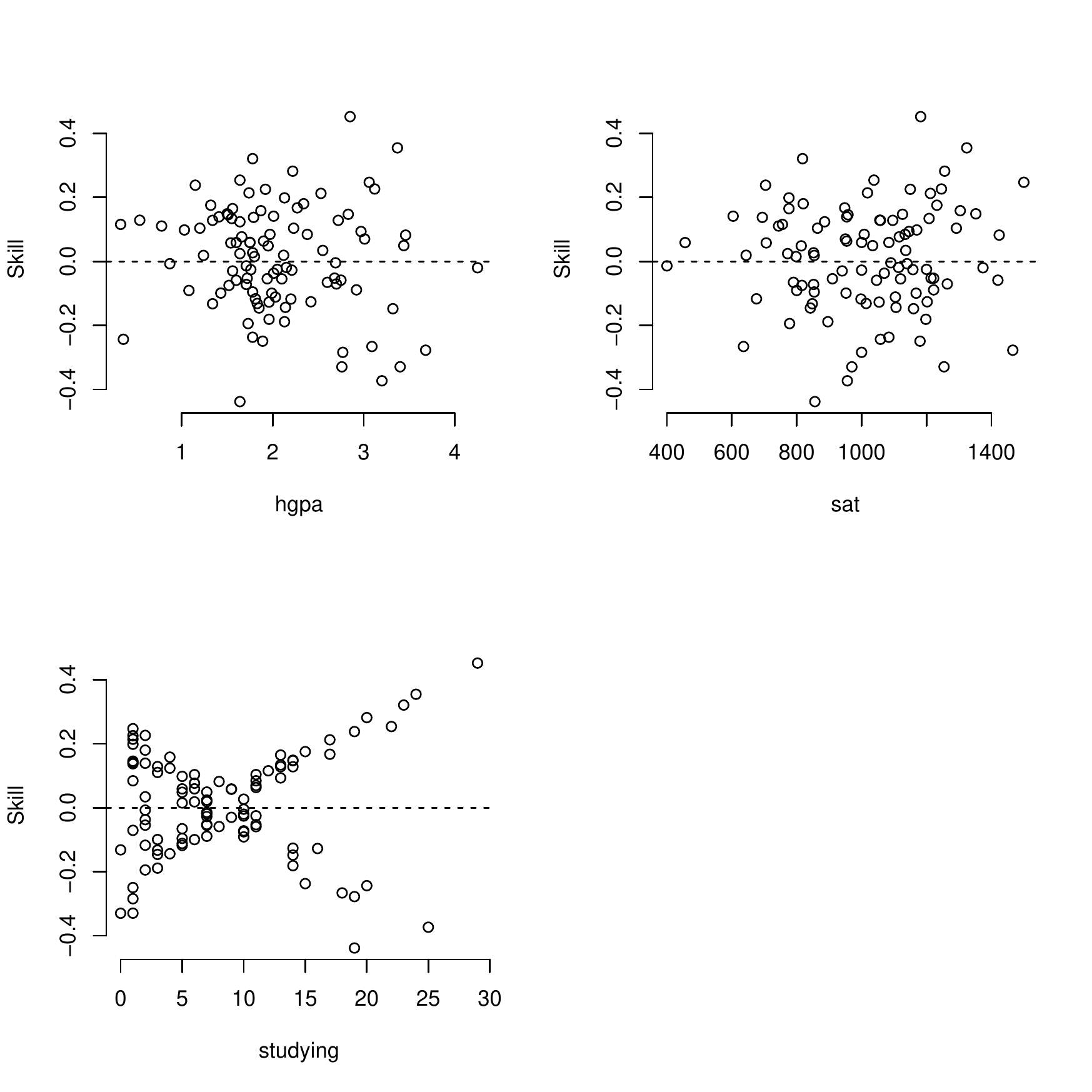}}
    \end{center}
\caption{\label{fig4} Top left: the skill plots for each observation of high school GPA. Top right: skill for each observation of SAT. Bottom left: skill for each observation of ``time studying".}
\end{figure}

The first graph in Fig. \ref{fig3} shows the CRPS calculated for each observation for the full and partial models; a one-to-one line is over-plotted. The addition of ``time studying" does not lead to uniform improvement.  The next graph are histograms of CRPS scores; the partial model is the dashed line. This is useful for a decision maker deciding how valuable either model is. Again, the best score is that which is related directly to the decisions made with the models. Whether CRPS is this score is situation dependent. The last graph shows the skill score for each observation of college GPA. A mixed bag, as far as ``time studying" goes. If we didn't already know, we'd suspect that ``time studying" is not adding much, and is even subtracting from, our knowledge of college GPA.

In Fig. \ref{fig4}  show the skill plots over each $X$ in the model for each observation. Skill is had for values greater than 0. The view that ``time studying" is nearly useless is confirmed. The last graph is particularly revealing. As ``time studying" moves to either extreme, the skill bifurcates, showing a process that ``can't make up its mind." If ``time studying" were truly valuable, the signal would be coherent.  At this point, conferring with the decision maker, the statistician might drop ``time studying" and compare his new ``full" model consisting only of high school GPA and SAT with (perhaps) a ``partial" model consisting only of an ``intercept". 

The verification process can be done on the already-observed data as an initial check on model goodness. The analogy is standard residual analysis. The same weaknesses apply, however, because the temptation to tweak the model to produce better verification measures will not be able to be resisted. Over-fitting and over-confidence will result, as always, but with an important twist. Since the model will be known and published in its predictive form, outsiders do not have to trust the in-sample verification. They can wait for new data and apply verification on them themselves. 

\section{Discussion}

Once importance or relevance are known, it is a mistake to say that $X$ is linked to, or is associated with, or predicts $Y$, or, worse, some variant of ``When $X$ equals $x$, $Y$ equals $y$". These are versions of a colossal misunderstanding, which is to say $X$ {\it causes} $Y$,  It is true that $X$ determines the uncertainty we have in $Y$, but {\it determines} is analogical; it has an ontological and epistemological sense. Probability is only concerned with the latter usage.  The only function probability has is to say how our assumptions $X$ determine epistemologically the uncertainty we have in $Y$. If we knew $X$ was a cause of $Y$, we have no need of probability. 

Importance and relevance are replacements for testing and estimation, but not painless ones. The recipient of an analysis is asked to do much more work than is usual in statistics. However, this is the more honest approach.  The benefit is that equation (\ref{third}) answers questions our customers ask us in the form they expect. The probabilities are in plain English and painless to interpret. Everything is stated in terms of observables. Everything is verifiable. The conditions on which the model relies are made explicit, made bare for all to see and to agree or disagree with.  Gone is the idea that there is one ``best" model which researchers have somehow discovered and which gives unambiguous results.  Gone also is the belief that the statistical analysis has proved a causal relationship.  

The model is made plain so that all can use it for themselves to verify predictions made with it. Everybody will be able to see for themselves just how useful the model really is. 

\bibliographystyle{apalike}
\bibliography{/home/briggs/projects/writing/logic}

\end{document}